\documentclass{webofc}

\usepackage[varg]{txfonts}   
\usepackage{hyperref}
\usepackage{float}
\usepackage{url}
\usepackage{}
\newcommand{\media}[1]{\left\langle #1 \right\rangle}

\hypersetup{colorlinks=true,citecolor=blue,urlcolor=blue,linkcolor=blue}

\begin{document}
%
\title{Probing the onset of collectivity via scaled particle spectra in ultrarelativistic nuclear collisions}
%
%

\author{\firstname{Thiago S.} \lastname{Domingues}\inst{1}\fnsep \and
        \firstname{Fernando G.} \lastname{Gardim}\inst{2}\fnsep\thanks{\email{fernando.gardim@unifal-mg.edu.br}} \and
        \firstname{David D.} \lastname{Chinellato}\inst{3}\fnsep \and
        \firstname{Gabriel S.} \lastname{Denicol}\inst{4}\fnsep \and
        \firstname{Andre V.} \lastname{Giannini}\inst{5}\fnsep \and
        \firstname{Matthew} \lastname{Luzum}\inst{1}\fnsep \and
        \firstname{Cicero D.} \lastname{Muncinelli}\inst{6}\fnsep \and
        \firstname{Jorge} \lastname{Noronha}\inst{7}\fnsep \and
        \firstname{Tiago Nunes} \lastname{da Silva}\inst{8}\fnsep \and
        \firstname{Jun} \lastname{Takahashi}\inst{6}\fnsep \and
        \firstname{Giorgio} \lastname{Torrieri}\inst{6}\fnsep    
}

\institute{
Instituto de F\'isica, Universidade de S\~ao Paulo, R. do Mat\~ao, 1371, S\~ao Paulo, SP, Brazil
\and 
Instituto de Ci\^encia e Tecnologia, Universidade Federal de Alfenas, Po\c cos de Caldas, MG, Brazil
\and
Stefan Meyer Institute for Subatomic Physics of the Austrian Academy of Sciences, Wiesingerstraße 4 1010 Vienna, Austria
\and
Instituto de F\'isica, Universidade Federal Fluminense,
Av. Milton Tavares de Souza, Niter\'oi, RJ, Brazil
\and
Faculdade de Ciências Exatas e Tecnologia, Universidade Federal da Grande Dourados, Dourados, MS, Brazil
\and
Universidade Estadual de Campinas, R. S\'ergio Buarque de Holanda, 777, Campinas, SP, Brazil 
\and
Illinois Center for Advanced Studies of the Universe \& Department of Physics, University of Illinois at Urbana-Champaign, Urbana, IL, 61801-3003, USA
\and
Departamento de F\'{i}sica, Centro de Ciências Físicas e Matemáticas, Universidade Federal de Santa Catarina, Campus Universit\'{a}rio Reitor Jo\~{a}o David Ferreira Lima, Florian\'{o}polis, Brazil\\
\hspace{14cm} (The ExTrEMe Collaboration)}
%
\abstract{We identify a novel scaling in the transverse momentum spectra of produced particles, obtained by removing the global scales of multiplicity and mean transverse momentum. Hydrodynamic simulations and experimental data reveal an almost universal scaled spectrum across centralities, systems, and even small systems, pointing to its origin in the collective, fluid-like dynamics of the QGP. Comparing this observable with Bayesian a priori distributions shows its independent constraining power on QCD transport properties, while also exposing limitations of current models. A detailed posterior analysis will be pursued in future work, opening a new avenue to refine our understanding of collectivity in heavy-ion collisions.
}
\maketitle
\vspace{-0.6cm}
\section{Introduction}
\label{intro}
Ultrarelativistic nuclear collisions create the Quark–Gluon Plasma (QGP), a nearly perfect fluid of quarks and gluons formed in heavy-ion collisions at the LHC~\cite{Busza:2018rrf}. Collective behavior has been established through flow observables $v_n$~\cite{JETSCAPE:2020mzn, Gardim:2012yp, Luzum:2013yya, ALICE:2016kpq}, but the information encoded in the full transverse momentum spectra, $dN/dp_T$, remains less understood. We introduce a new observable, the scaled spectra,
\begin{equation}
U(x_T) \equiv \frac{\media{p_T}}{N}\frac{dN}{dp_T}=\frac{1}{N}\frac{dN}{dx_T},
\label{eq:universal}
\end{equation}
with $x_T=p_T/\media{p_T}$, obtained by removing global scales of multiplicity $N$ and mean transverse momentum $\media{p_T}$. Remarkably, the scaled spectra collapse onto a universal curve across centralities and systems (Pb–Pb, Xe–Xe, and p–Pb), breaking only at high $p_T$ and in very small systems (p–p)~\cite{Muncinelli:2024izj}. Hydrodynamic models reproduce the scaling event-by-event, establishing it as a robust signature of collective fluid dynamics. 
In such an event-by-event analysis our observable becomes related to the so-called $v_0$ coefficient \cite{v0ref,v0ref1}, with the scaling signaling that this coefficient does not vary with centrality and energy.

While Bayesian analyses have so far relied mainly on integrated observables such as $v_n$, the strong fluid connection of this scaling makes the $p_T$ spectra a natural ingredient, providing independent constraints on QCD transport properties. As the onset and breakdown of hydrodynamic behavior remain open problems, the universal scaling offers a powerful tool to experimentally pinpoint the limits of collectivity in hadronic collisions.

\section{Universality of scaled spectra in hydro and experimental data}
\label{sec-2}
Fluid-dynamical models have been widely used to describe QGP in heavy-ion collisions. Recently, Ref.\cite{Muncinelli:2024izj} showed that, in event-by-event hybrid simulations\footnote{Hybrid simulations combine initial conditions with a pre-equilibrium stage, 2+1D viscous hydrodynamics, particlization, and hadronic scattering}, the scaled spectra of pions, Eq.\eqref{eq:universal}, becomes independent of centrality once the system, energy, and model parameters are fixed, combining the full spectral information into a single distribution.

\begin{figure}[H]
\centering
\vspace{-.1cm}\includegraphics[width=.85\textwidth]{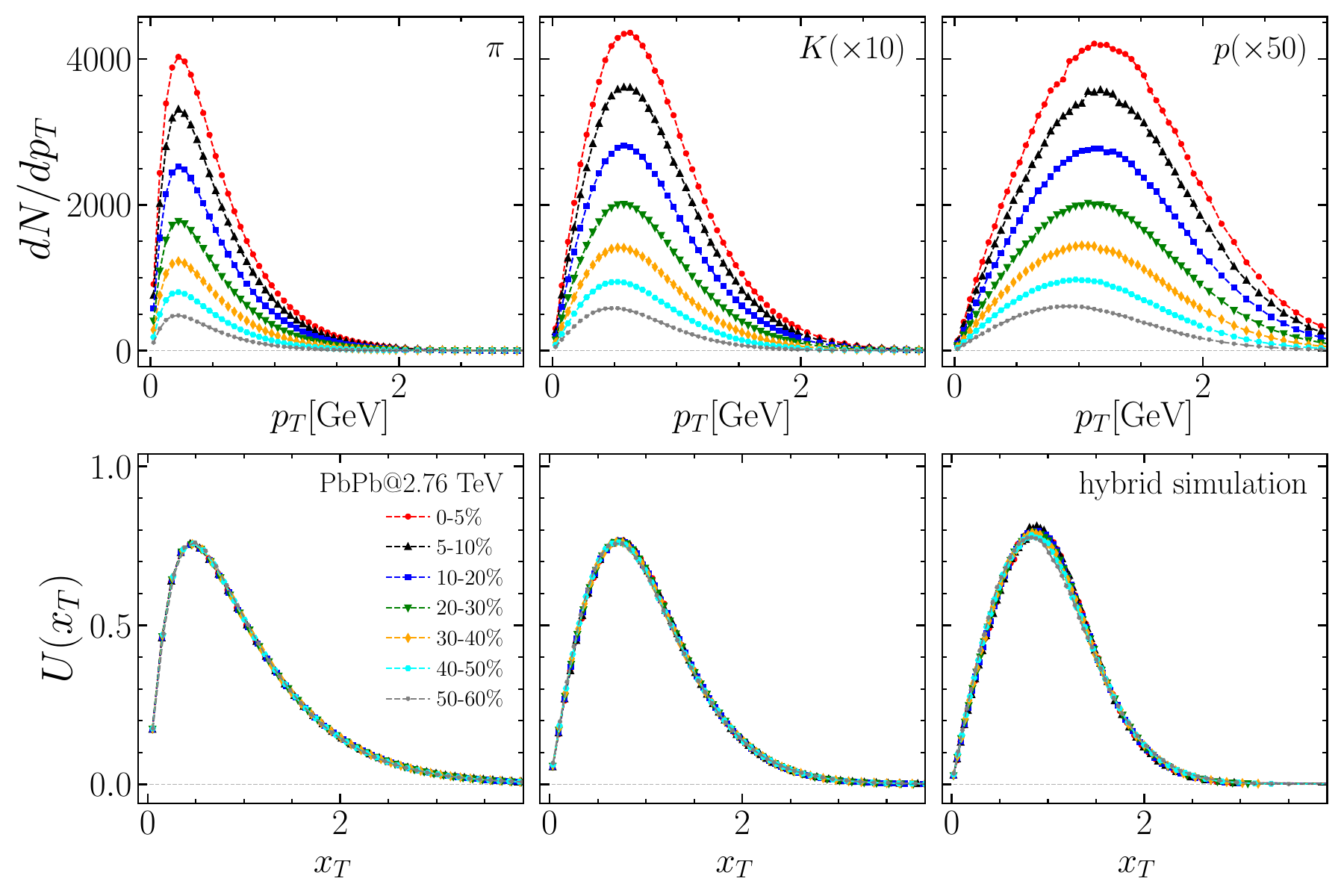}
\vspace{-.3cm}\caption{First row: particle spectra of pions, kaons and protons in different centralities, obtained from hybrid simulation. The spectra for kaons was scaled by a factor of 10, and for protons by a factor of 50. Second row: the scaled spectra for pions, kaons and protons, obtained from de particle spectra.}
\label{fig:pikp}       
\end{figure}

To extend this result beyond pions, we present hybrid simulation results for pions, kaons, and protons, using parameters obtained from Bayesian analyses of Pb–Pb collisions at 2.76 TeV and Au–Au at 0.2 TeV, employing the \textit{maximum a posteriori} (MAP) set of the Grad viscous correction~\cite{JETSCAPE:2020mzn}. Figure~\ref{fig:pikp} shows results from 5,000 simulations across seven centrality bins. The first row shows the mean $p_T$ spectra for each species, with clear centrality dependence. The second row shows the scaled spectra $U(x_T)$ obtained via Eq.~\eqref{eq:universal}. While pion and kaon spectra collapse onto a universal curve, proton spectra exhibit slight 
variations as a function of centrality.

\begin{figure}[h]
\centering
\includegraphics[width=.85\textwidth,clip]{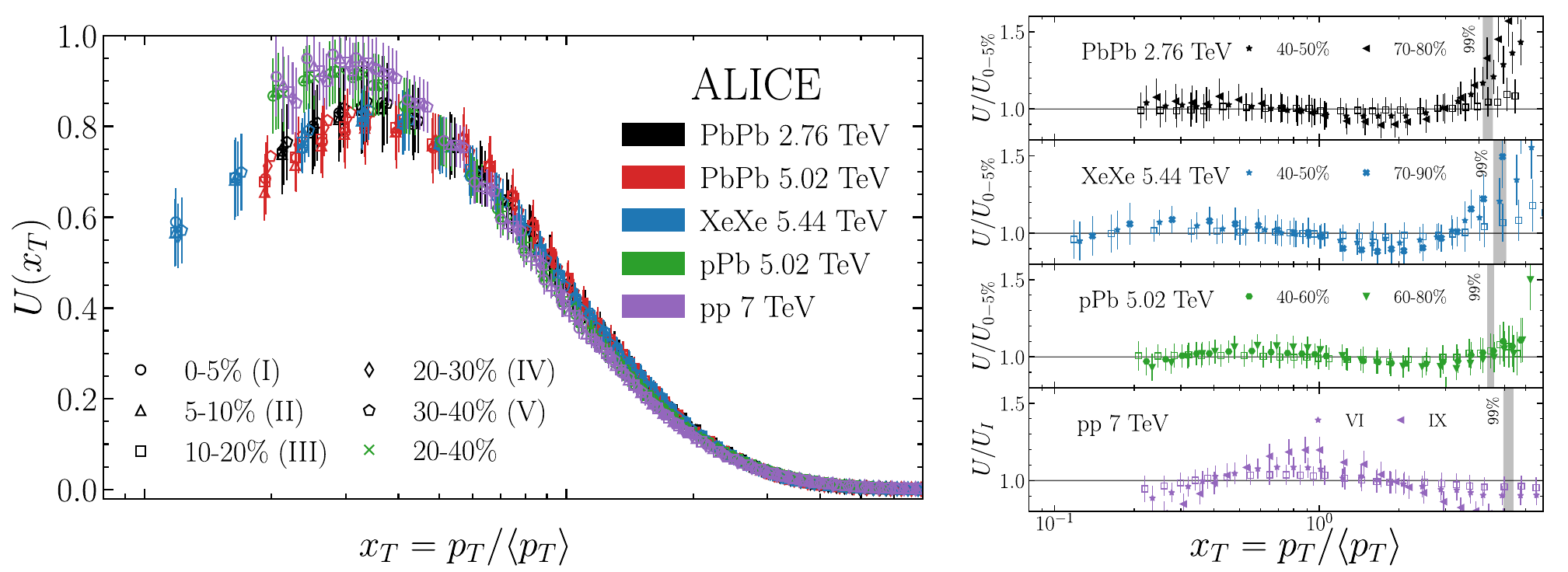}
\vspace{-.3cm}\caption{(Left) The scaled spectra $U(x_T)$ for pions in Pb-Pb at 2.76 TeV \cite{pb276ALICE:2013mez}, Xe-Xe 5.44 TeV \cite{ALICE:2021lsv}, p-Pb 5.02 TeV \cite{ppb502ALICE:2013wgn}, and p-p 7 TeV \cite{pp7ALICE:2018pal} from experimental data. The ratio between $U$ coming from four centralities and $U$ at 0-5\%. 99$\%$ of the pions are produced to the left of the vertical gray lines.}
\label{fig:alice}       
\end{figure}

We now present the scaled spectra for pions in experimental data across different systems and energies. Unlike our simulations, the data include additional physical effects, such as jets. Figure~\ref{fig:alice} (right) shows $U(x_T)$ with uncertainties propagated from $p_T$-spectra, $\media{p_T}$, and $N$. For fixed large system and energy, $U(x_T)$ collapses onto a nearly universal curve across centralities. In p–Pb, universality is observed but the spectrum differs from heavy-ion collisions. Smaller uncertainties in $\media{p_T}$ are needed for a conclusive statement. The ratio to central (0–5\%) spectra confirms universality in Pb–Pb and Xe–Xe, though deviations grow in peripheral events, suggesting the onset of non-hydrodynamic physics. At large $p_T$, scaling breaks down as expected from the transition to (semi-)hard processes.

Results for small systems show that p–Pb exhibits scaling across centralities within uncertainties, while in p–p the scaling clearly breaks down except, possibly, in high-multiplicity events.  Thus, the novel scaling phenomena unveiled in this study gives further evidence for hydrodynamic
behavior in p-Pb collisions, and possibly in high-multiplicity p-p collisions, providing information about collectivity in small systems independently from anisotropic flow signatures.
\vspace{-.7cm}\section{Incorporating scaled spectra into Bayesian analysis}
\label{sec:bayesian_analysis}
The scaling behavior of $U(x_T)$ emerges as a distinctive signature of the hydrodynamic evolution of the QGP and should be included in global analyses, beyond $N$ and $\langle p_T\rangle$ from $dN/dp_T$. The transport coefficients of the QGP can be constrained through model-to-data comparisons employing Bayesian parameter estimation. Using the model and prior parameter space of the JETSCAPE Collaboration~\cite{JETSCAPE:2020mzn}, we extracted universal spectra for pions and performed a comparison based exclusively on $U(x_T)$ within the Grad viscous correction model.

\begin{figure}[h]
\centering
\includegraphics[width=.95\textwidth,clip]{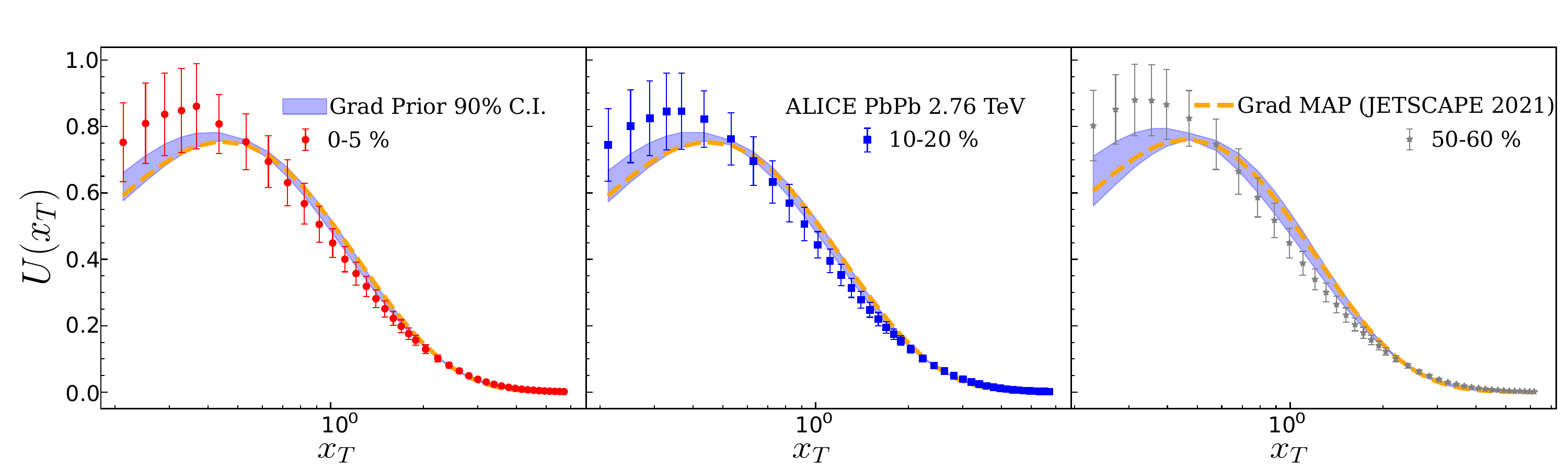}
\vspace{-.3cm}\caption{Prior predictions for the Grad prescription compared to ALICE measurements in Pb–Pb collisions at $\sqrt{s_{\rm NN}}=2.76$ TeV for 0–5\%, 10–20\% and 50–60\%. The 90\% credible intervals, light blue bands, are the prior predictions. MAP prediction from JETSCAPE~\cite{JETSCAPE:2020mzn} is shown in orange dashed line.}\vspace{-.5cm}
\label{fig:prior_Grad}      
\end{figure}

Trained Gaussian emulators using the JETSCAPE simulational data allow us to compare a priori distribution predictions of the Grad viscous correction model with the experimental $U(x_T)$ in Fig.~\ref{fig:prior_Grad}, where the parameters 
have been varied over the entire prior space of
Ref.~\cite{JETSCAPE:2020mzn}.
We observe that the scaled spectra remains universal,
even over a large variation in parameters, and there
is limited flexibility to tune the model to this observable.
Moreover, results obtained with the JETSCAPE MAP parameter set (calibrated on $p_T$-integrated observables), shown as the orange dashed line in Fig.~\ref{fig:prior_Grad}, are inconsistent with experimental data. This comparison suggests that current state-of-the-art models may not capture the observed universal shape. A detailed Bayesian analysis will be presented in future work.

In conclusion, this work reveals a remarkable universality in the transverse momentum spectra of ultrarelativistic nuclear collisions, observed both in fluid-dynamical model and experimental data once the spectra are rescaled by $N$ and $\media{p_T}$, applying to pions, kaons and protons. The presence of this scaling in heavy-ion collisions, and its breakdown at high transverse momentum, highlights its connection to the hydrodynamic regime of QCD matter. Using ALICE data, we demonstrated that the scaling holds for Pb–Pb, Xe–Xe, and p–Pb collisions, but not for p–p, providing new evidence for collective dynamics and the possible formation of a QGP in small systems.

The scaled spectrum emerges as a robust observable to investigate the onset and limits of hydrodynamic behavior across different systems and energies. By training emulators to predict $U(x_T)$ using a state-of-the-art model, we demonstrated that this observable is not well described by current models, highlighting its independent constraining power on the transport properties of QCD matter, thereby complementing traditional analyses based on $p_T$-integrated observables and anisotropic flow. These results motivate further theoretical and experimental studies of this scaling phenomenon, to uncover the interplay between collective and non-collective dynamics encoded in the $p_T$ spectrum, expanding conventional particle physics studies, and improving our understanding of the QGP.\\

%
\vspace{-.2cm}\noindent \textit{Acknowledgments:} F.G.G.~was supported by CNPq (Conselho Nacional de Desenvolvimento Cientifico) through 307806/2025-1 and FAPEMIG APQ-00544-23. F.G.G. and T.N.dS, are supported by CNPq 409029/2021-1. F.G.G., M.L., T. N.dS, G.S.D and T.S.D are supported by CNPq through the INCT-FNA 312932/2018-9. T.S.D. was supported through a PhD fellowship 141432/2025-0. C.D.M.~was supported through FAPESP grant 2025/01122-0. We also acknowledge support from FAPESP project 2023/13749-1.
%

\vspace{-.2cm}\bibliography{reff}
\end{document}